\documentclass{aa}
\usepackage{graphicx}
\usepackage{natbib}
\usepackage{epsfig}
\def\ut#1{\mathop{\vtop{\ialign{##\crcr
     $\hfil\displaystyle{#1}\hfil$\crcr\noalign
     {\kern1pt\nointerlineskip}\hbox{$\hfil\sim\hfil$}\crcr
     \noalign{\kern1pt}}}}}

\def\undersymbol#1#2{\mathop{\vtop{\ialign{##\crcr
     $\hfil\displaystyle{#2}\hfil$\crcr\noalign
     {\kern1pt\nointerlineskip}\hbox{$\hfil#1\hfil$}\crcr
     \noalign{\kern1pt}}}}}

\begin{document}

  \title{A characterization of the NGC 4051 soft X-ray spectrum as observed by {\it XMM}-Newton}
      \author{A.A. Nucita\inst{1}, M. Guainazzi\inst{1}, A.L. Longinotti\inst{2}, M. Santos-Lleo\inst{1}, Y. Maruccia\inst{3}, \and S. Bianchi\inst{4}}
              \institute{XMM-Newton Science Operations Centre, ESAC, ESA, PO Box 78, 28691 Villanueva de la $\rm Ca\tilde{n}ada$, Madrid, Spain \and
              MIT Kavli Institute for Astrophysics and Space Research 77 Massachusetts Avenue, NE80-6011 Cambridge, MA 02139\and
               Dipartimento di Fisica, Universit\`a del Salento, and {\it INFN}, Sezione di Lecce, CP 193, I-73100 Lecce, Italy\and
              Dipartimento di Fisica, Universit\`a degli Studi Roma Tre, via della Vasca Navale 84, 00146 Roma, Italy }

   \offprints{A. A. Nucita, \email{anucita@sciops.esa.int}}
   \date{Submitted: XXX; Accepted: XXX}
{
  \abstract
   {Soft X-ray high resolution spectroscopy of obscured AGNs shows a complex soft $X$-ray
   spectrum dominated by emission lines of He and H-like transitions of elements from carbon to neon, as well as L-shell transitions due to iron ions.}
{In this paper we characterize the {\it XMM}-Newton RGS spectrum of the Seyfert 1 galaxy NGC 4051 observed during a low flux state and infer the physical properties of the emitting and absorbing gas in the soft $X$-ray regime.}
{X-ray high-resolution spectroscopy offers a powerful diagnostic tool because the observed spectral features strongly depend on the physical properties of matter (ionization parameter $U$, electron density $n_e$, hydrogen column density $N_H$), which in turn are tightly related to the location and size of the $X$-ray emitting clouds. We carried out a phenomenological study to
identify the atomic transitions detected in the spectra. This study suggests that the spectrum is dominated by emission from a photoionized plasma. Then we used the photoionization code Cloudy to produce synthetic models for the emission line component and the warm absorber observed during phases of high intrinsic luminosity.}
    {The low state spectrum cannot be described by a single photoionization component. A multi-ionization phase gas with an ionization parameter in the range of $\log U \sim 0.63-1.90$ and a column density $\log N_H = 22.10-22.72$ cm$^{-2}$ is required, while the electron density $n_e$ remains unconstrained. A warm absorber medium is required by the fit with the parameters $\log U \sim 0.85$, $\log N_H = 23.40$ and $\log n_e \ut< 5$. The model is consistent with an $X$-ray emitting region at a distance $\ut> 5\times 10^{-2}$ pc from the central engine.
}
   {}
}


   \authorrunning{Nucita et al.}
   \titlerunning{The soft X-ray spectrum of NGC 4051}
   \maketitle
%

\section{Introduction}

It is commonly accepted that the center of active galaxies (Active Galactic Nuclei -AGNs) hosts a massive black hole (with a mass in the range $10^6-10^9$ M$_{\odot}$) accreting the surrounding material via the formation of a disk. How the energy released from the central engine interacts with the local environment and contributes to the history of the host galaxy is one of the crucial question of present astrophysical research.
In this respect, while the mechanisms of energy output in the form of radiation and relativistic jets are quite well understood, it also seems that the outflowing winds have an important role in the overall energy budget.
Although the origin of these winds is still controversial, at our present level of understanding the narrow-line regions, the inner part of an obscuring torus (\citealt{blustin2005}) and the black hole accretion disk \citep{elvis2000} are all possible locations.

X-ray obscured AGNs (with an intrinsic column density $N_H\ut> 10^{22}$ cm$^{-2}$) are not completely dark in the soft X-ray band. High resolution {\it XMM}-Newton and Chandra observations revealed a complex spectrum dominated by emission lines from He-and H-like transitions of elements from carbon to neon as well as by L-shell transitions of {Fe\,\textsc{xvii}} to {Fe\,\textsc{xxi}} ions (\citealt{sakoa}, \citealt{ali}, \citealt{sambruna}, \citealt{armentrout2007}). This gas, which shows the signature of a photoionization process (\citealt{ali}, \citealt{bianchi2007}), is sometimes referred to as a warm mirror.

In unobscured AGNs a modification of the output energy spectrum may also occur as a consequence of absorption by a warm ionized gas along the line of sight. The properties of these so called warm absorbers can be summarized as follows: i) average ionization parameter in the range $\log \xi =0-3$, ii) total column density in the range $\log N_H = 21-22$ cm$^{-2}$, iii) outflow velocities of hundred of km s$^{-1}$ (see e.g. \citealt{blustin2005}, but also \citealt{steenbrugge}). Evidence of a multi-phase warm absorber gas was also recently reported for Mrk~841 (\citealt{longinotti2009}).

In general, detecting warm mirror signatures is easier in sources in low flux states, because the emission features are not outshone by the continuum radiation. This was the case for the Seyfert 1 galaxy Mrk 335, whose soft X-ray spectrum resembled the spectra of obscured AGNs  when the source was observed at low state (\citealt{longinotti2008}), but does not show any evidence of a warm absorber in the high flux state (\citealt{longinotti2007}).

The overall properties of the warm mirror (even if it is poorly constrained) and the warm absorber (as described above) are similar so that there is the possibility that they represent the same physical system.
Conversely, the interplay between the warm absorber and warm mirror regimes is best studied in sources that display both components.

The source NGC 4051, a narrow-line Seyfert galaxy at the redshift of $0.00234$, was at the center of many past investigations in the X-ray band because it offers a unique laboratory where to test present theories and models about the physics of AGNs. The X-ray emission is characterized by rapid variations \citep{lamer2003,ponti2006} sometimes showing periods of low activity (see \citealt{lawrence1987} and \citealt{uttley}). Its power spectral distribution (PSD) in high state resembles the behavior of a galactic black hole system (\citealt{mchardy}). At high X-ray flux, the spectrum of the galaxy is characterized by a power law with photon index $\Gamma\sim 1.8-2$ which becomes harder above 7 keV where a reflection component from cold matter has been observed. On long time-scales, the X-ray light curve of NGC 4051 shows low state flux periods of several months during which the spectrum in the energy range 2-10 keV becomes harder ($\Gamma \simeq 1$) and shows a strong iron $K\alpha$ line (as found by \citealt{guainazzi98} in Beppo-SAX data). A soft X-ray excess is also evident.

As reported by \citet{warmabsorber}, the high state X-ray spectrum of NGC 4051 in the soft band is a combination of continuum and emission line components. Curvature in the spectrum cannot be explained with simple models, i.e. a single power law or a black body, because an ionized absorber-emitter has to be taken into account as well. In this context, \citet{krongold} showed that the evolution in time of the properties of the warm absorber can constrain the physical parameters of the absorbing gas. In  particular they find that at least two different ionization components are required with matter densities of $\simeq 10^6$ cm$^{-3}$ and $\ut> 10^7$ cm$^{-3}$, thus placing the warm absorber in the vicinity of the accretion-disk. Dynamical arguments permit us to infer that the warm absorber gas originates in a radiation-driven high-velocity outflow in accretion disk instabilities (\citealt{krongold}).

On the other hand, as shown by \cite{pounds}, the low state flux spectrum of NGC 4051 is dominated by narrow emission lines and radiative recombination continua (RRC) from hydrogenic and He-like carbon, oxygen, neon and nitrogen. To be specific, a fit to the identified RRCs yields a mean temperature for the emitting gas of $T\simeq 4\times 10^{4}$ K, which favors a scenario invoking a photoionization process. In this case, the soft X-ray spectrum of NFG 4051 in low state is similar to that observed for the prototype Seyfert 2 galaxy NGC 1068 (see \citealt{ali}).

Below we do not repeat the analysis of the EPIC data but refer to \citet{pounds} for more details on the main results obtained in the energy band 0.3-10 keV. We only say that a comparison between the EPIC PN data for the 2001 and 2002 observations shows that the high state observation flux level is a factor $\sim 5$ greater with respect to the low state. Furthermore, the spectrum shows a gradual flattening of the continuum slope from 3 keV up to 6.4 keV. It was also noted that when the fit to the 0.3-10 keV band continuum is extrapolated down in the soft X-ray (0.3-3 keV) a strong excess appears in both the two observations, and as is clear from the RGS spectrum, it can be explained by a blending effect of fine structures (emission lines).

Here we first conducted a phenomenological study of the emission lines identified in the spectrum of NGC 4051 and compare our results with those known in literature.
We further compared the RGS emission line spectrum with synthetic spectra generated with the photoionization code Cloudy 8 (\citealt{ferland}).
For this purpose we followed a similar approach as in \cite{armentrout2007} (to which we refer for more details) on NGC 4151.

The paper is structured as follows: in Sect. 2 we briefly describe the reduction of the {\it XMM}-Newton data set and describe our
phenomenological analysis of the soft $X$-ray spectrum of NGC 4051. In Sects. 3 and 4 we give details on the Cloudy model developed and address some conclusions.
\begin{figure*}[htbp]
\vspace{9.5cm} \includegraphics{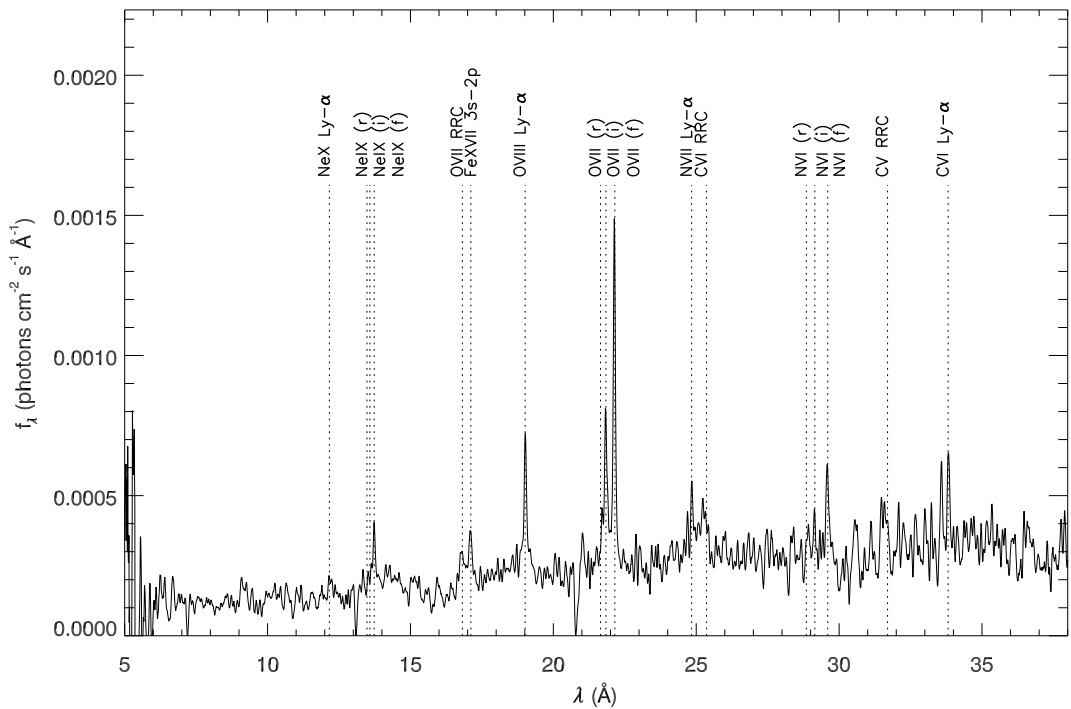}
\caption{Fluxed RGS spectrum of NGC 4051 (low state). The first order spectra of the two RGS cameras were combined and the resulting spectrum smoothed with a triangular kernel. The identified lines are labeled with the corresponding ion transition name and vertical dashed lines (the big dips at $\simeq 13$ \AA and $\simeq 21$ \AA in this plot are due to CCD gaps).}
\label{spectrum}
\end{figure*}

\section{A phenomenological study of the low state of NGC 4051: data reduction and line identification}

The source NGC 4051 ($\alpha=12^{h}03^{m}09.6^{s}$ and $\delta=44^{d}31^{m}53.0^{s}$) was observed by the {\it XMM}-Newton satellite on two occasions: on May 2001 for $\simeq 122$ ks and on November 2002 for $\simeq 52$ ks. While the former observation coincided with a period where the central engine was bright (with luminosity of $7\times10^{41}$ erg s$^{-1}$ in the $0.3-10$ keV, \citealt{pounds}), the latter corresponded to a low state X-ray flux (corresponding to a luminosity of $1.5\times10^{41}$ erg s$^{-1}$ in the $0.3-10$ keV, \citealt{pounds}) due to a low nuclear activity. This observation was conducted $\sim 20$ days after the onset of the low state (\citealt{pounds}).
Below we focus on the low state data analysis, because the warm absorber observed at high state was already well studied with physical models (\citealt{warmabsorber}, \citealt{krongold}, \citealt{steenbrugge}). On the contrary, no attempt has yet been made to model the warm mirror in the low state with a self consistent physical model.

The ODF files (OM, MOS, PN and RGS) were processed with the {\it XMM}-Science Analysis System (SAS version $8$). Hence the raw data were reduced using SAS tasks with standard settings and the most update calibration files to produce the source and background spectra as well as the corresponding response matrices for the RGS cameras.

We used XSPEC 12.5.1 (\citealt{arnaud}) for our quantitative analysis and adopted the cosmological parameters $H_0=70$ km s$^{-1}$ Mpc$^{-1}$, $\Omega_{\Lambda}=0.73$ and $\Omega_{m}=0.27$.

To study the soft X-ray spectra of NGC 4051 in more detail, we then examined the first order spectra obtained by the {\it XMM}-Newton gratings. The spectral resolution of RGS in the first order spectrum is FWHM=72 m\AA~and the calibration  in wavelength is accurate up to 8 m\AA~corresponding to FWHM$\simeq 620$ km s$^{-1}$ and $\Delta v\simeq 69$ km s$^{-1}$ at 35 \AA (\citealt{xmmuserbook}). Below we use the unbinned RGS 1 and RGS 2 spectra for the quantitative analysis. In Fig. \ref{spectrum}, we show the fluxed RGS spectrum in the wavelength range 5-38 \AA.

As can be clearly seen, the 2002 RGS spectrum of NGC 4051 shows an unresolved continuum with a predominance of emission lines (see for comparison the RGS spectrum of NGC 1068 presented in \citealt{ali}). By contrast, the high state RGS spectra of the same source show a higher continuum flux level with a pronounced curvature around $\sim 15$ \AA  and several absorption features \citep{pounds} typical of a warm absorber. Interestingly, the {N\,\textsc{vi}}, {O\,\textsc{vii}} and {Ne\,\textsc{ix}} forbidden lines are seen in both observations with the same flux level (\citealt{pounds}).

The phenomenological spectral analysis follows the {\it local fits} method described in \citet{bianchi2007}. In particular, the unbinned spectra are divided in intervals of $\simeq 100$ channels wide and Gaussian profiles are used to account for all identified emission lines, with the line centroid energy as the only free parameter of the fit.

Analogously, free-bound transitions (i.e. Radiative Recombination Continua) were also modeled as Gaussian profiles with free line width. Best-fit values of these widths are reported in Table \ref{completarrc} together with their errors. The local continuum was modeled as a power law with a fixed photon index $\Gamma = 1$ and free normalization. For line triplets and for emission lines close to free-bound transitions, the relative distance between the central energies was frozen to the value predicted by atomic physics.

We used the C-statistic as the estimator of the goodness of the performed fit (\citealt{cash}). For any line under investigation, the emission feature was considered {\it detected} if, when repeating the fit without any Gaussian profile, we obtained a value of the C-statistic which differed from the previous one by at least 2.3, corresponding to the 68$\%$ confidence level (or, equivalently, 1 $\sigma$ for one interesting parameter) \citep{arnaud}.

The results of the phenomenological fit to the emission lines are reported in Table \ref{completa}. We recall that the fluxes of the identified emission lines were estimated integrating over the Gaussian line profiles with which the emission features are modeled. This operation resulted in flux values somehow higher (up to a factor of $\sim 1.5$) than those reported in \citet{pounds} even if the corresponding measured equivalent widths are fully consistent with the values quoted in the above mentioned paper. The origin of the discrepancy remains unknown.

Here we give the best-fit parameters for the transitions identified in the soft X-ray spectrum of NGC 4051.
For those lines that were not identified (i.e. $\Delta C <$ 2.3), but which are part still of triplet features, we give the upper limits to the corresponding flux value.
We detected no evidence for either inflows or outflows, with an upper limit on the velocity of $\sim 200$ km s$^{-1}$, consistent with that estimated by \cite{pounds}.
This was calculated from the full width half maximum of the distribution of residual velocity as derived from the difference between the expected
and the measured laboratory wavelengths (see Fig. \ref{redshift} and also Table \ref{completa}).
\begin{table*}
\begin{center}
\begin{tabular}{|l|c|c|c|c|l|}
\hline
Line ID & $\lambda_{exp} ({\rm {\AA}})$& $\lambda_{obs} ({\rm {\AA}})$ & ${\rm v}~({\rm km s^{-1}})$ & Flux ($\times 10^{-14}$ cgs) & $\Delta {\rm C}$ \\
\hline

{Ne\,\textsc{x}} Ly-$\alpha$   ~&$12.134$~&$12.13_{-0.13}^{+0.03}$~&$-23_{-3100}^{+ 800}$ ~&$1.5_{-1.4}^{+1.6}$~&3 \\
{Ne\,\textsc{ix}} (r)          ~&$13.447$~&$13.45_{-0.02}^{+0.02}$~&$-7_{- 360}^{+ 400}$  ~&$\ut< 0.81$       ~&-- \\
{Ne\,\textsc{ix}} (i)          ~&$13.550$~&$13.55_{-0.02}^{+0.02}$~&$23_{- 360}^{+ 400}$  ~&$\ut< 0.77$       ~&-- \\
{Ne\,\textsc{ix}} (f)          ~&$13.700$~&$13.70_{-0.02}^{+0.02}$~&$-40_{- 360}^{+ 400}$ ~&$4.7_{-1.2}^{+1.3}$~&50 \\
{Fe\,\textsc{xvii}} 3s-2p      ~&$17.073$~&$17.06_{-0.02}^{+0.04}$~&$-135_{- 380}^{+ 710}$~&$3.3_{-1.5}^{+1.4}$~&21 \\
{O\,\textsc{viii}} Ly-$\alpha$ ~&$18.969$~&$18.98_{-0.02}^{+0.01}$~&$150_{- 310}^{+ 190}$ ~&$7.0_{-1.1}^{+1.2}$~&135 \\
{O\,\textsc{vii}} (r)          ~&$21.600$~&$21.58_{-0.01}^{+0.01}$~&$-202_{-210}^{+ 150}$ ~&$\ut< 0.80$~&-- \\
{O\,\textsc{vii}} (i)          ~&$21.790$~&$21.80_{-0.20}^{+0.25}$~&$194_{-2380}^{+3400}$ ~&$7.2_{-1.6}^{+1.7}$~&74 \\
{O\,\textsc{vii}} (f)          ~&$22.101$~&$22.10_{-0.02}^{+0.03}$~&$32_{-250}^{+ 400}$   ~&$15.0_{-1.9}^{+2.1}$~&238 \\
{N\,\textsc{vii}} Ly-$\alpha$  ~&$24.781$~&$24.79_{-0.02}^{+0.03}$~&$198_{-220}^{+ 330}$  ~&$2.0_{-0.8}^{+0.9}$ ~&17 \\
{N\,\textsc{vi}} (r)           ~&$28.787$~&$28.78_{-0.01}^{+0.02}$~&$-104_{-150}^{+220}$  ~&$\ut< 0.92$         ~&-- \\
{N\,\textsc{vi}} (i)           ~&$29.083$~&$29.07_{-0.01}^{+0.02}$~&$-94_{-150}^{+220}$   ~&$1.3_{-0.6}^{+0.5}$ ~&6 \\
{N\,\textsc{vi}} (f)           ~&$29.534$~&$29.52_{-0.01}^{+0.02}$~&$-103_{-150}^{+220}$  ~&$3.6_{-0.8}^{+0.7}$ ~&38 \\
{C\,\textsc{vi}} Ly-$\alpha$   ~&$33.736$~&$33.76_{-0.02}^{+0.03}$~&$233_{-140}^{+ 240}$  ~&$3.5_{-1.1}^{+1.3}$ ~&29 \\
\hline
\end{tabular}
\end{center}
 \caption{Best-fit parameters for the transitions identified in the soft X-ray spectrum of NGC 4051. From left to right we give the name of the detected ion transition, the expected centroid wavelength (\AA) in the rest-frame as extracted from the CHIANTI database \citep{chianti}, the observed wavelength (\AA), the corresponding shift in velocity (km s$^{-1}$), the line flux (in units of $10^{-14}$ erg s$^{-1}$ cm$^{-2}$) and  the associated difference in C-statistics, i.e. $\Delta C$. All lines that were not identified ($\Delta C < 2.3$) but are part of triplet features are also reported.}
\label{completa}
\end{table*}
\begin{table*}
\begin{center}
\begin{tabular}{|l|l|c|c|c|l|}
\hline
Line ID & $\lambda_{exp} ({\rm {\AA}})$ & $\lambda_{obs} ({\rm {\AA}})$ & ${\rm v}~({\rm km s^{-1}})$ & Flux ($\times 10^{-14}$ cgs) & $\Delta {\rm C}$ \\
& & & & & \\
\hline
{O\,\textsc{vii}}~ RRC    ~& $16.771$ &$16.78_{-0.07}^{+0.04}$~&~~$220_{-1200}^{+800}$~& $ 3.8_{-1.2}^{+1.3}$~&    38 \\
{C\,\textsc{v}}~ RRC    ~& $31.622$ &$31.51_{-0.06}^{+0.10}$~&$-1070_{-600}^{+800}$~& $ 3.4_{-2.0}^{+2.0}$~&   9 \\
{C\,\textsc{vi}}~ RRC    ~& $25.304$ &$25.19_{-0.04}^{+0.10}$~&$-1250_{-500}^{+900}$~& $ 6.4_{-3.0}^{+16.0}$~&  24 \\
\hline
\end{tabular}
\end{center}
\caption{The same as in Table \ref{completa} but for the identified RRCs.}
\label{completarrc}
\end{table*}

\subsection{Radiative recombination continua (RRC)}
The electron temperature $T_e$ can be inferred by studying the profiles of the radiative recombination continua (RRC). In the RGS spectrum of NGC 4051 the RRCs detected with $\Delta C \ut > 2.3$ correspond to {O\,\textsc{vii}}, {C\,\textsc{v}} and {C\,\textsc{vi}} (see Table \ref{completarrc} for details). Expressing the temperature widths in eV to $kT_e$ (\citealt{lied}), we estimate them to be
\begin{equation}
\begin{array}{ll}
T_{O~VII}=\Big(5.0^{+4.0}_{-2.0}\Big)\times 10^{4}~{\rm K}~,\\ \\
T_{C~VI}=\Big(3.0^{+3.1}_{-1.3}\Big)\times 10^{4}~{\rm K},\\ \\
T_{C~V}=\Big(1.2^{+0.8}_{-0.8}\Big)\times 10^{4}~{\rm K},
\end{array}
\end{equation}
respectively, so that the average gas temperature is $T_e = \Big(3.1^{+2.5}_{-1.6}\Big)\times 10^{4}~{\rm K}$, which agrees well with the result quoted by \citet{pounds}. It is to note that the low temperature found in this way is an indication that collisional ionization and excitation processes are negligible (\citealt{lied}).

In this phenomenological analysis, we used Gaussian profiles to fit the RRC features, which are in principle asymmetric. Still we verified that the use of a more appropriate model, as e.g. {\it redge} in XSPEC, gives consistent results.

\subsection{He-like triplet diagnostic}
We detected the most intense lines of He-like ions in the range 5-35 \AA. The transitions between the $n=2$ shell and the $n=1$ ground state shell as the resonance line (${\bf r:~}~1s^2$ $^1S_0$-$1s2p$ $^1P_1$), the two inter-combination lines (${\bf i:~}~1s$$^2$$^1S_0$-$1s2p$ $^3P_{2,1}$, often blended) and the forbidden line (${\bf f:~}$~$1s^2$$^1S_0$-$1s2s$$^3S_1$) were detected. As demonstrated by \citet{porquet} the relative emission strength of the r, i and f lines are good indicators of the physical conditions of density and temperature of the gas. Using standard notation we defined the ratios $R=f/i$, $L=r/i$ and $G=(f+i)/r$ (\citealt{porter2007}). Figure \ref{ovii} shows the triplet of the {O\,\textsc{vii}} complex (forbidden, inter-combination and resonance lines) locally fitted by a power law and three Gaussian. In this case, following the phenomenological fit approach described in the previous Section, we only had a measurement for the fluxes of the $f$ and $i$ components (see Table \ref{completa}).

With the flux measurements quoted in Table \ref{completa}, the previous relations give $R=2.1^{+0.5}_{-0.6}$, $L=0.07\pm 0.06$ and $G= 47_{-39}^{+34}$. Analogously, for the NVI triplet we get $R=2.7_{-1.9}^{+1.2}$, $L\ut<0.69$ and $G\ut>5.4$ (poorly constrained because we only got an upper limit to the $r$ line flux value), respectively. For {Ne\,\textsc{ix}} we had a lower limit only on the $R(\ut> 5.6)$ ratio, while the ratio of the {O\,\textsc{viii}} Ly-$\alpha$ to the {O\,\textsc{vii}} forbidden intensity lines results in $0.47_{-0.13}^{+0.14}$. These line ratios are consistent with the results  by \citet{pounds}.

\subsection{Results of the phenomenological study: evidence of photoionized gas}

The results obtained from the phenomenological study allow us to highlight some considerations on  the physical conditions of the $X$-ray emitting gas in NGC 4051. Indeed, according to the study of \citet{porquet}, a value of the G ratio higher than 4 is a strong indication of a photoionized gas. An estimate of the gas electron density $n_e$ can be done when the other two line ratios L and R are taken into account. In the particular case of the {O\,\textsc{vii}} triplet line ratios quoted above, the electron density is constrained to be  $n_e \ut< 10^{10}$ cm$^{-3}$ for a pure photoionized gas (\citealt{porquet}). Note however that the line intensities obtained from the phenomenological study described above do not account for a warm absorber, which is not taken into account in the model.

An additional constraint on the electron density value can be obtained noting that the {\it XMM}-Newton observation of NGC 4051 in its low state occurred $\sim 20$ days after the source entered this regime.
Because the {O\,\textsc{vii}} triplet line intensity is consistent with that measured during high flux states (\citealt{pounds}), it is believed that the recombination time of the {O\,\textsc{vii}}
is larger than 20 days,thus implying (for a gas temperature of $\simeq 10^4$ K) a more stringent constraint on the electron density of $n_e \ut< 10^5$ $cm^{-3}$ (\citealt{pounds}).
\begin{figure}[h]
\vspace{7.5cm} \includegraphics{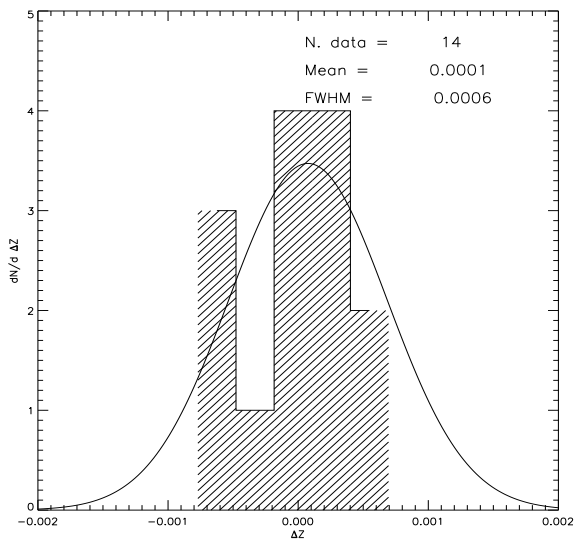}
\caption{Distribution of residual velocity for all lines in Table \ref{completa} with respect to the cosmological one. Note that the observed shifts are consistent with the cosmological ones, i.e. no outflow or inflow is observed. The solid line represents the Gaussian best-fit to the data.}
\label{redshift}
\end{figure}
\begin{figure}[htbp]
\vspace{6.0cm} \includegraphics{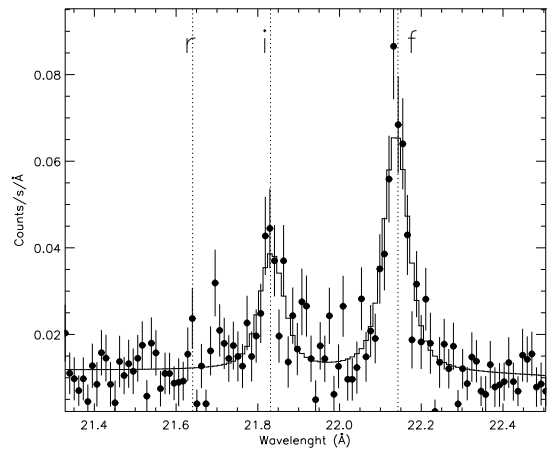}
\caption{Zoom around the {O\,\textsc{vii}} triplet lines in the low flux state. The emission lines correspond to the f, i and r components, respectively. Only an upper limit to the resonance line r can be obtained (see text). The solid line represents the best-fit obtained with a power law + Gaussian lines model (details in text).}
\label{ovii}
\end{figure}
\section{Fitting the spectra with the photoionization code Cloudy}

\subsection{General properties of the model}
The results of the phenomenological analysis in Sect. 2 show that the bulk of the spectrum, measured by the RGS during the $X$-ray low state in NGC 4051, is dominated by photoionization as already suggested
by other authors (see e.g. \citealt{pounds}).

In this section we use the photoionization code Cloudy (\citealt{ferland}) for modeling the overall spectrum of NGC 4051 in its low state assuming a plane parallel geometry with the central engine shining on the inner face of the cloud with a flux density depending on the ionization parameter $U$.

The spectrum produced by a photoionized nebula critically depends on the spectral energy distribution (SED) of the ionizing continuum. Below we adopt  the AGN SED as in \citet{korista}. In a typical AGN  the observed continuum can be well represented by a SED characterized by several components: a big blue bump with temperature  $T_{BB}\simeq 10^{6}~K$ (1~Ryd), a power law with a low energy exponential cut-off in the infrared region at $kT_{IR}=0.01$ ~Ryd; in the $X$-ray band ($1.36$ eV-100 keV) the SED is well approximated by a power law with an exponential cut-off for energies lower than 1 Ryd; finally, for energies greater than 100 keV an exponential fall as  $\propto \nu^{-2}$ is usually assumed . We also included in the modeling the cosmic microwave background so that the incident continuum has a non-zero intensity for long wavelengths. Hence the AGN spectrum is described by the law
\begin{equation} \label{sed}
F(\nu)=\nu^{\alpha_{UV}}e^{-\frac{h\nu}{kT_{BB}}}e^{-\frac{kT_{IR}}{h\nu}}+A\nu^{\alpha_{x}}e^{-\frac{1~{\rm Ryd}}{E({\rm Ryd})}}~,
\end{equation}
where $\alpha_{UV}\simeq -0.50$, $\alpha_{x}$ is the spectral photon index and the constant $A$ is obtained requiring that
$F(2 ~{\rm keV})/F(2500 ~{\rm \AA})\simeq 403.3^{\alpha_{ox}}$ \citep{korista1997}, where $F(2 ~{\rm keV})$ and $F(2500~ {\rm \AA})$ are the flux densities at 2 keV and 2500 \AA, respectively.

To determine $\alpha _x$ and $\alpha _{ox}$, we used the Epic data corresponding to the NGC 4051 high state observation. The resulting 0.2-10 keV energy band spectrum was fitted with a photoelectrically absorbed power law model within XSPEC, thus allowing us to measure $\alpha _x= 0.96\pm 0.05$ and $F(2 ~{\rm keV})=(2.91\pm0.01)\times 10^{-29}$ erg s$^{-1}$ cm$^{-2}$ Hz$^{-1}$, respectively.

From the OM instrument we estimated the aperture photometry of the target in the UVM2 filter (centered at $2310$ \AA) obtaining a flux density of $F(2310 ~{\rm \AA})=(1.50\pm0.01)\times 10^{-14}$ erg s$^{-1}$ cm$^{-2}$ \AA$^{-1}$. From the flux densities at 2 keV and 2500 \AA, the X-UV flux density ratio results in $\alpha _{ox}\simeq -1.14$.

Once the AGN SED (erg s$^{-1}$ cm$^{-2}$ Hz$^{-1}$) is known, it is straightforward to show that the number of hydrogen-ionizing photons $Q$\footnote{The default energy range used by the Cloudy code to evaluate the number of ionizing photons $Q$ is $1$ Ryd - $7.354\times 10^{6}$ Ryd, \citep{ferland}.}, the electron density $n_e$ and the dimensionless ionization parameter $U$ are related by $U=Q/4 \pi r^2 n_e c$ with $r$ the distance between the central engine and the innermost illuminated layer of the clouds. Here we require that integrating over the SED (between 13.6 eV and 13.6 keV) we get the ionizing luminosity $L_{ion}\simeq 4.1\times 10^{42}$  erg s$^{-1}$ \citep{warmabsorber}. In Fig. \ref{sed} we compare the SED used in this paper (solid line) with that given in \citet{warmabsorber}. We recall that the dimensionless ionization parameter $U$ does not depend on the flux below $13.6$ eV. The two spectral energy distributions give rise to comparable integrated fluxes in the 0.3-10 keV energy band (within a few percent).

Note also that through the well known definitions of the ionizing luminosity $L_{ion}$, of the number of ionizing photons $Q$ (\citealt{ferland}) and the used SED $F(\nu)$, it is possible to estimate a useful conversion relation between the dimensionless ionization parameter $U$ and the ionization parameter $\xi$ as given in \citet{tarter}, i.e. $\xi\simeq 20 U $ erg cm s$^{-1}$.

\begin{figure}[htbp]
\vspace{6.0cm} \includegraphics{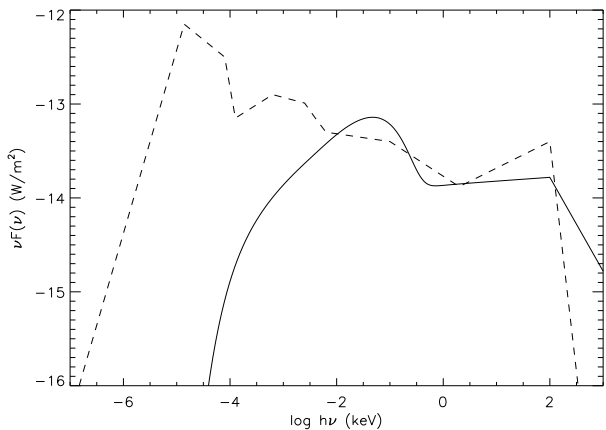}
\caption{SED used in the paper (solid line). For comparison we also give (dashed line) the SED given in \citet{warmabsorber}. This was taken from the NED database and is likely to contain contamination from the hosting galaxy.}
\label{sed}
\end{figure}

\subsection{Fit to the RGS spectrum}
Assuming the standard AGN continuum described above, we generated a grid of reflected spectra from a photoionized nebula, varying the ionization parameter $\log U$, the electron density $\log n_e$ and the total column density $\log N_H$. The free parameters spanned the ranges $\log U$=[-1.0,4.0]; $\log (n_e/{\rm cm^{3}})=[2,12]$ and $\log (N_H/{\rm cm^{2}})$=[19,24] in steps of 0.1 dex, respectively.

Initially we extracted the line intensities from these simulations for all lines detected in the soft spectrum of NGC 4051. Hence, we tried to determine the best model that can describe the RGS line spectrum in the whole $5-35$ \AA band. Following the
procedure described in \citet{longinotti2008}, we calculated the value of the merit function for each grid model
\begin{equation}
\chi ^2 =\sum \frac{(I_c-I_o)^2}{\sigma_o ^2}~,
\end{equation}
where $I_o$ is the intensity of each of the identified lines (with statistical
error  $\sigma_o$) and $I_c$ is the intensity as predicted by Cloudy,
both normalized to the value of the {O\,\textsc{viii}} Ly-$\alpha$ line.

Minimizing the merit function quoted above gives a best-fit model ($\chi^2 \simeq 13.8$ with degrees of freedom $\nu=10$) corresponding to the parameter values $\log U=0.4$, $\log (n_e/{\rm cm^{3}})=4.4$ and $\log (N_H/{\rm cm^{2}})=21.8$.
A quantitative measure of the fit goodness for the used model is given by the Chi-square Probability Function $Q(\chi^2,\nu)$ as defined in \citet{press}. If the single phase component model is
the true representation of the data, the probability to obtain the observed $\chi^2$ value is as high as $Q\simeq 12\%$. In this case, the model consisting of a single ionization state can be statistically rejected.

We therefore investigated more complex models, including an additional warm mirror and one warm absorber covering the combination of emitting components. For this approach to be fruitful the constraints provided by the continuum shape are crucial. Below we will fit the whole RGS spectrum globally.

\subsection{Global fit to the RGS spectrum}
We generated additive and multiplicative fits tables (with the same grid of parameters as before) to account for both the emission and absorption features observed in the RGS spectrum, and imported them within XSPEC as described in \citet{porter}. Our final model can be described by the formula $ phabs*mtab(n_e,N_H,U)*\{\sum atab(n_e,N_H,U)\}$.
Here,  {\it mtab} and {\it atab} indicate the warm absorber component and the reflected component part of the spectrum depending on the electron density, hydrogen column density and ionization parameter, respectively. In the model, the redshift of each component is fixed to the cosmological value due to the lack of measurable velocity shifts from the phenomenological analysis (Fig. \ref{redshift}), while all the other parameters are free to vary. In the fit procedure we fixed the column density of neutral hydrogen to the average value observed in the Galaxy along the line of sight to NGC 4051, i.e. $1.32\times10^{20}$ cm$^{-2}$ (\citealt{dickey}).

The fit does not formally depend on values of $n_e\ut< 10^{9}$ cm$^{-3}$, which is expected because the ratios of the He-like triplets are insensitive to the electron density in this region of the space parameter (\cite{porquet}). Given the constraint on this parameter derived from the source time variability, we fixed its value to $10^{5}$ cm$^{-3}$ hereafter.

\begin{figure}[htbp]
\vspace{6.5cm} \includegraphics{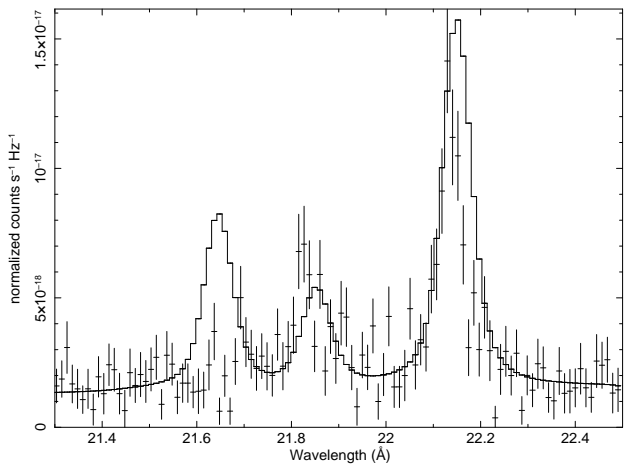}
\caption{Comparison of a model with two emission components (solid line) with the observed RGS spectrum (zoom around the {O\,\textsc{vii}} triplet).}
\label{fig6}
\end{figure}

We recursively increased the number of Cloudy additive components until this operation resulted in a statistically significant improvement of the fit quality.
We found that two emission and one absorption components are required to fit the data. In particular,
the final model corresponds to a value of the C-statistic of 6300 with 5178 d.o.f. and the model parameters
are given in Table 4. Conversely, when the warm absorber component is not taken into account the fit visibly
worsens and converges to a C-statistic value of 9452 with 5182 d.o.f. In this case,
the line intensities corresponding to the He-like transitions are not correctly estimated, with specifically
the recombination line of the {O\,\textsc{vii}} triplet well over-estimated (see Fig. \ref{fig6}).

\begin{figure}[htbp]
\vspace{8.0cm} \includegraphics{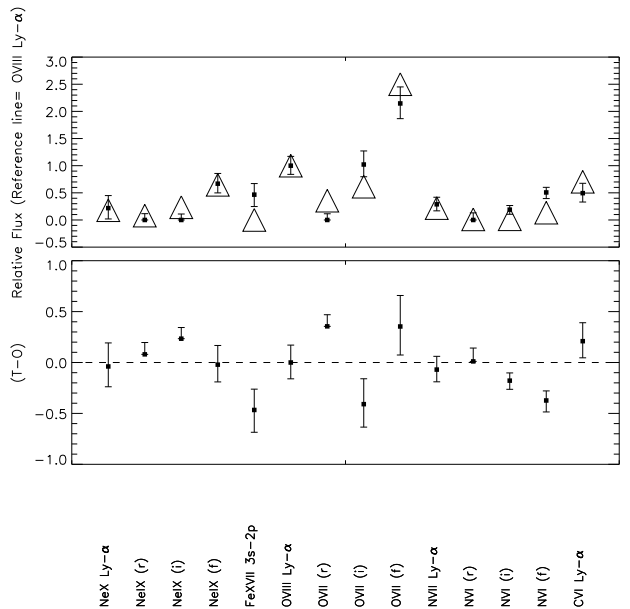}
\caption{Intensities of the observed (filled squares) and simulated (triangles) lines corresponding to the best-fit and normalized to the {O\,\textsc{viii}} Ly-$\alpha$ flux (see also Table \ref{completa}). Residuals are in the right bottom panel.}
\label{fig4}
\end{figure}

We then extracted the line fluxes predicted by the best-fit model and compared them with the observations.
In Figure \ref{fig4}, we show with filled squares the intensities of all observed (see Table \ref{completa})
and simulated (triangles) lines once normalized to the {O\,\textsc{viii}} Ly-$\alpha$ flux. The lower panel of the same figure
shows the residuals between observation and theory. Note that the simulation underestimates the
contribution of the {Fe\,\textsc{xvii}} transitions, because the Cloudy database is inaccurate for the corresponding
atomic parameters (see e.g. \citealt{bianchi2010}). The normalized intensities of the observed lines as well
as the Cloudy predictions are also reported in Table \ref{table3} for clarity with the missing value of the {Fe\,\textsc{xvii}} transition.

\begin{table*}
\begin{center}
\begin{tabular}{|l|l|c|c|}
\hline
Line ID & $\lambda_{exp} ({\rm {\AA}})$& Observed ratio & Cloudy predicted ratio \\
\hline
{Ne\,\textsc{x}} Ly-$\alpha$   ~&$12.134$~&$0.2_{-0.2}^{+0.2}$~&$ 0.2$\\
{Ne\,\textsc{ix}} (r)          ~&$13.447$~&$ \ut<0.1$~&$ 0.1$ \\
{Ne\,\textsc{ix}} (i)          ~&$13.550$~&$ \ut<0.1$~&$0.2$ \\
{Ne\,\textsc{ix}} (f)          ~&$13.700$~&$ 0.7_{-     0.2}^{+     0.2}$~&$0.6$ \\
{Fe\,\textsc{xvii}} 3s-2p      ~&$17.073$~&$ 0.5_{-0.2}^{+0.2}$~&$--$ \\
{O\,\textsc{viii}} Ly-$\alpha$ ~&$18.969$~&$ 1.0_{-0.2}^{+0.2}$~&$1.0$ \\
{O\,\textsc{vii}} (r)          ~&$21.600$~&$ \ut< 0.1$~&$0.3$ \\
{O\,\textsc{vii}} (i)          ~&$21.790$~&$ 1.0_{-     0.2}^{+     0.2}$~&$0.6$ \\
{O\,\textsc{vii}} (f)          ~&$22.101$~&$ 2.2_{-0.3}^{+0.3}$~&$2.5$ \\
{N\,\textsc{vii}} Ly-$\alpha$  ~&$24.781$~&$ 0.3_{-0.1}^{+0.1}$~&$0.2$ \\
{N\,\textsc{vi}} (r)           ~&$28.787$~&$ \ut<0.13$~&$0.01$ \\
{N\,\textsc{vi}} (i)           ~&$29.083$~&$ 0.1_{-    0.1}^{+    0.1}$~&$0.01$ \\
{N\,\textsc{vi}}(f)           ~&$29.534$~&$ 0.5_{-     0.1}^{+    0.1}$~&$0.1$ \\
{C\,\textsc{vi}} Ly-$\alpha$   ~&$33.736$~&$ 0.5_{-0.2}^{+0.2}$~&$0.7$ \\
\hline
\end{tabular}
\end{center}
\caption{Intensities of the observed lines and of those simulated by Cloudy corresponding to the best-fit and normalized to the {O\,\textsc{viii}} Ly-$\alpha$ flux (see text and Fig. \ref{fig4} for more details).}
\label{table3}
\end{table*}

\begin{figure*}[h]
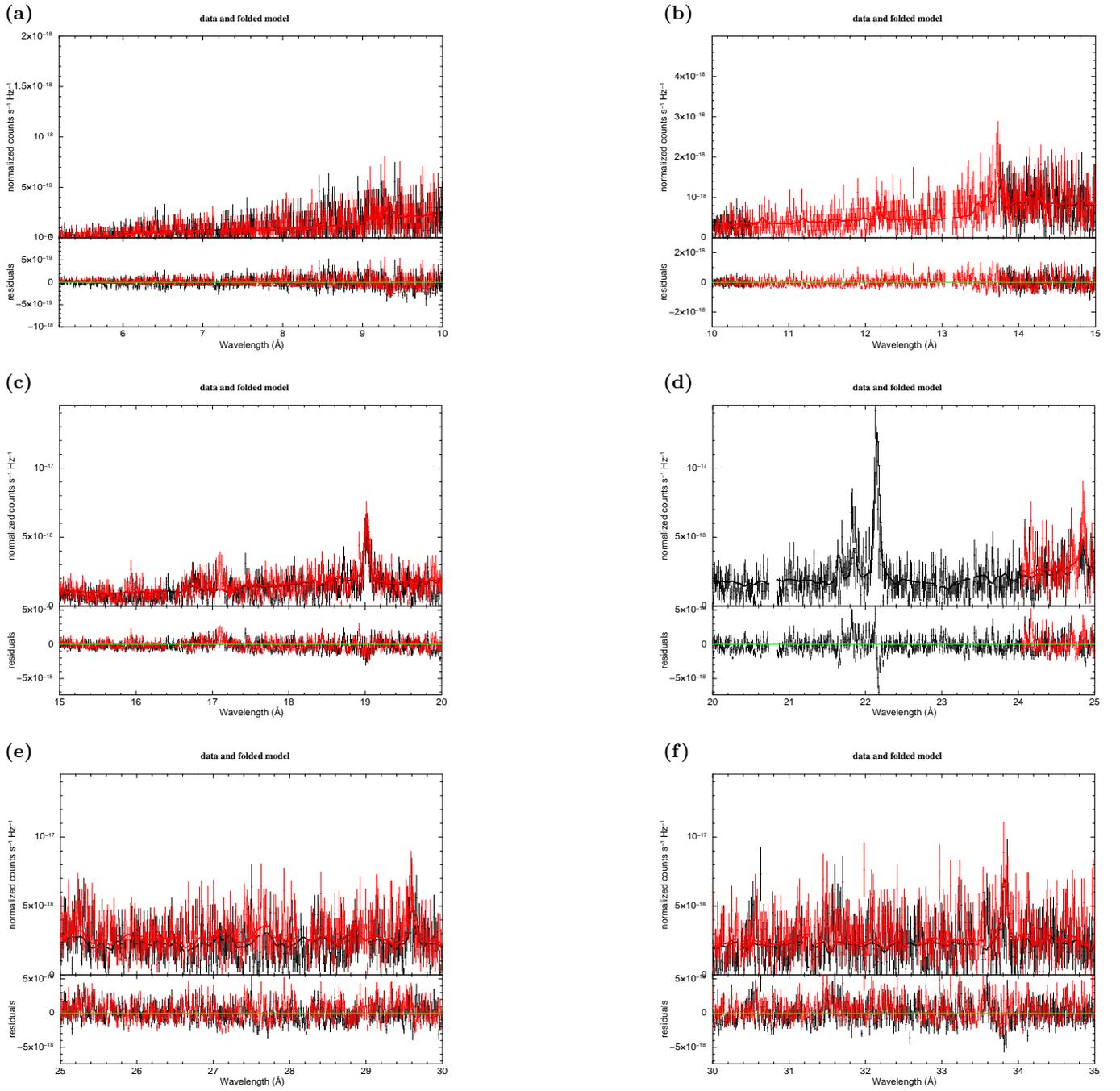

\begin{center}
$\begin{array}{c@{\hspace{1in}}c}
\multicolumn{1}{l}{\mbox{\bf (a)}} &
	\multicolumn{1}{l}{\mbox{\bf (b)}} \\ [-0.53cm]
\includegraphics[width=6cm, angle=-90]{5-10_new.ps}&
\includegraphics[width=6cm,angle=-90]{10-15_new.ps}\\ [0.8cm]
\multicolumn{1}{l}{\mbox{\bf (c)}} &
\multicolumn{1}{l}{\mbox{\bf (d)}} \\ [-0.53cm]
\includegraphics[width=6cm,angle=-90]{15-20_new.ps}&
\includegraphics[width=6cm,angle=-90]{20-25_new.ps}\\ [0.8cm]
\multicolumn{1}{l}{\mbox{\bf (e)}} &
	\multicolumn{1}{l}{\mbox{\bf (f)}} \\ [-0.53cm]
\includegraphics[width=6cm,angle=-90]{25-30_new.ps}&
\includegraphics[width=6cm,angle=-90]{30-35_new.ps}\\ [0.8cm]
\end{array}$
\end{center}
\caption{The 5-35 \AA~RGS 1 (red data points) and RGS 2 (black data points) spectra are shown together with the best-fit Cloudy model described in the text and corresponding to the physical parameters reported in Table 4.}
\label{spectrum_all}
\end{figure*}

In Table 4 we give the relevant quantities estimated from the fit procedure (i.e. $n_e$, $N_H$ and $U$) together with their respective errors at the 90$\%$ confidence level for one interesting parameter. The 5-35 \AA~low-state spectrum of NGC 4051 is plotted in Fig. \ref{spectrum_all} with the best-fit Cloudy model superimposed on the observed RGS 1 (red) and RGS 2 (black) data and residuals in the lower part of each panel.

Note that to estimate the covering factor of the source, we extracted the luminosity of the most prominent emission line, {O\,\textsc{vii}} (f), as predicted by Cloudy. For the two reflection components included in our model, we simulated the expected spectrum the SED described in Sect. 3.1 and fixing the electron density $n_e$ and hydrogen column density $N_H$ to the best-fit values given in Table 4. In addition, the SED was normalized to the ionizing luminosity $L_{ion}\simeq 4.1\times 10^{42}$  erg s$^{-1}$ \citep{warmabsorber}. Hence we fixed the distance from the source to the inner shell of the cloud to the lower values reported in Table 4 for each of the reflection components, i.e. $d\simeq 0.22$ pc and $d\simeq 0.05$ pc for the {\it low} and {\it high} component, respectively. After we defined the luminosity of the source, the emission line luminosities were predicted by Cloudy (see e.g. \citealt{hazy1}).

With a redshift of 0.00234, the expected total intensity of the {O\,\textsc{vii}} (f) line is $\simeq 10.6\times 10^{-13}$ erg s$^{-1}$ cm$^{-2}$. Assuming a filling factor of unity, the ratio of the observed {O\,\textsc{vii}} (f) intensity (see Table \ref{completa}) to the Cloudy expected value gives an estimate of the covering factor value, which turns out to be $\simeq 0.14$.

\begin{table*}
\begin{center}
\begin{tabular}{|c|c|c|c|c|c|}
\hline
  &  {\bf $\log U$} & {\bf $\log (n_e/{\rm cm^{3}})$}& {\bf $\log (N_H/{\rm cm^{2}})$}& {\bf $d$} (pc) & {\bf $\Delta_{f=1}$} (pc)\\
  &              &     &     &       &          \\
\hline
 {\bf Low}      &  $0.63_{-0.03}^{+0.05}$ & $\ut< 5$ & $22.10_{-0.04}^{+0.30}$ & $\ut> 0.22$& $\ut> 0.04$\\
 {\bf High}     &  $1.90_{-0.10}^{+0.20}$ & $\ut< 5$ & $22.72_{-0.13}^{+0.25}$ & $\ut> 0.05$& $\ut> 0.20$\\
\hline
 {\bf Warm Abs.}      &  $0.85_{-0.02}^{+0.02}$ & $\ut< 7$ & $23.36_{-0.01}^{+0.01}$ & $\ut> 0.02$& $\ut> 0.01$\\
\hline
\end{tabular}
\caption{Best-fit parameters for the three components of the adopted Cloudy model used to fit the RGS data. Errors are given at the 90\% confidence level for one interesting parameter. Columns are: the ionization parameter, electron density, Hydrogen column density, distance from the central ionizing source, and the average $X$-ray emitting source size evaluated as $\Delta\simeq N_H/f n_e$, where $f$ is the filling factor. Here, for simplicity we have assumed $f \simeq 1$.}
\end{center}
\label{tabellafinale}
\end{table*}

\section{Discussion}
Most of the information on the physics and geometry of gas in AGNs is inferred by means of optical spectroscopy and imaging techniques with which it was shown that the AGN central high energy emission is the main source of ionizing photons with an occasional contribution from collisionally ionized plasma. In the last years X-rays observations acquired an important role in AGN studies particularly since Chandra showed the existence, at least for Seyfert 2 galaxies, of extended (a few kpc) X-ray emission \citep{bianchi2006} similarly to what was observed in the optical band.
High resolution spectroscopy in the soft X-ray band ($0.2-2$ keV) confirms the overall scenario, and photoionization seems to be
the dominant ionization mechanism which results in a spectrum characterized by recombination lines from He- and H-like transitions of C to Si elements and by Fe-L transitions. In this respect, X-ray high-resolution spectroscopy offers a powerful diagnostic tool because the observed spectral features strongly depend on the physical properties of matter (ionization parameter $U$, electron density $n_e$, hydrogen column density $N_H$ as well as size and location of the emitting clouds).

The Seyfert 1 object NGC 4051 shows a very rich emission line $X$-ray spectrum when observed in low-flux state. According to the analysis we conducted on the {\it XMM}-Newton RGS data, the observed soft $X$-ray features originate in a low-density photoionized gas.
In order to constrain the physical properties of the photoionized gas, we simulated synthetic spectra via the Cloudy software (\citealt{ferland}) and compared them to the RGS data with standard minimization techniques. We found that to describe the overall soft $X$-ray spectrum, at least a three-phase gas is required (two emission components and one warm absorbing component). Referring to the emission components respectively as {\it low} and {\it high} ionization components, our fit procedure gave us their physical properties. For the {\it low} component we have $\log U\simeq 0.63$, and $\log (N_H/{\rm cm^{2}})\simeq 22.10$ and for the {\it high} component we have $\log U\simeq 1.90$, and $\log (N_H/{\rm cm^{2}})\simeq 22.20$. Using Cloudy we get for the electron density $n_e$ an upper limit of $\log (n_e/{\rm cm^{3}}) \simeq 9$, which reduces to $\log (n_e/{\rm cm^{3}}) \simeq 5$ when the recombination time scale of {O\,\textsc{vii}} is taken into account. Even if the warm absorber gas seems to be required by our fit procedure, its parameters are poorly constrained. Thus it is characterized by $\log U\simeq 0.85$, $\log (N_H/{\rm cm^{2}})\simeq 23.36$, and $\log (n_e/{\rm cm^{3}}) \ut< 7$.

This technique was successfully applied before to the Seyfert 1 Mrk~335 and the Seyfert 2/starburst galaxy NGC~1365  (\citealt{longinotti2008,guainazzi2009}).
The main difference is that NGC~4051 is characterized by a strong warm absorber component in the high flux state that is still affecting the spectrum even when the nuclear flux is attenuated. Indeed, we found out in our analysis of the low flux state data that the effect of the line of sight medium is not negligible, particularly not in the modeling of the resonance line of the {O\,\textsc{vii}} triplet (see Fig. \ref{fig6}) which is close to several absorption features. For example, the resonance line could be weakened by the same line in absorption\footnote{As noted by \citet{sakob} and \citet{ali}, the resonance line of the {O\,\textsc{vii}} triplet could be also enhanced by photoexcitation. Note however that this would also result in a boost of all the higher order resonance transitions of the H-like and He-like ions (Ly-$\beta$, Ly-$\delta$, He-$\beta$ and He-$\delta$), but this enhancement is not currently observed.} (see e.g. \citealt{krongold}). Nonetheless, the physical parameters of the warm absorber cannot be well-constrained by the analysis of the low flux state data (see Table 4).

The average distance $r$ of each of the photoionized plasma-emitting components from the nuclear source can be estimated by the definition of the ionization parameter $U$ after normalizing to the ionization luminosity $L_{ion}$. However, our results are insensitive to values of the electron density $n_e$ lower than $10^5$ cm$^{-3}$. In this limit, we can only determine a lower limit of the $X$-ray-emitting gas location (Table 4).

The  analysis carried out in this paper allowed us to identify two ionization states for the line emitting gas and one warm absorber medium. It is interesting to note that \\

- The $X$-ray emitting region can be placed at a distance of $r\ut> 0.05$ pc.

Indeed, \citet{warmabsorber} found that the NGC 4051 $X$-ray narrow-line regions can be placed at a distance of the same order of magnitude. This was also confirmed by the Chandra ACIS-S images of the same galaxy (\citealt{uttley}), which showed a size of the diffuse emission smaller than that of the optical narrow-line regions (30 -220 pc, \citealt{christopoulou1997}), thus implying a clear separation between the $X$-ray and optical emissions.

This is also naturally expected as a consequence of projection effects: as shown by \citet{schmitt2003}, who studied a sample of 60 Seyfert galaxies with the Hubble Space Telescope, the Seyfert 1 narrow-line regions objects are more circular and compact than those in the Seyfert 2 galaxies, with the Seyfert 2 subsample characterized by more elongated shapes. This agrees well with the unified picture according to which the conical narrow-line region of a Seyfert 1 galaxy is observed close to the axis of symmetry, while that of a Seyfert 2 galaxy is observed from an orthogonal line of sight.

Furthermore, the scale-length found in this paper is consistent with the inner radius of the torus in NGC~4051 as determined by \citet{blustin2005}, i.e. $r\simeq 0.15$~pc.\\

- The NGC 4051 low state warm absorber is poorly constrained but its existence is nevertheless required by the fit. In particular, we found a lower limit of the warm absorber distance $\simeq 0.02$ pc, i.e. at least a factor $10$ larger than that measured in the high state flux (\citealt{krongold}).

Indeed, by using the long $XMM$-Newton exposure of NGC 4051 in its high flux state and studying the time evolution of the ionization states of the $X$-ray absorbers, \citet{krongold} were able to put severe constraints on the physical and geometrical properties of the warm absorber medium.

They specifically found that the warm absorber consists of two different ionization components which are located within $3.5$ lt-days (or $0.0029$ pc) from the central massive black hole. This result allowed the authors to exclude an origin in the dusty obscuring torus because the expected dust sublimation radius\footnote{The torus inner edge has to be at a distance larger than the dust sublimation radius $r_{sub}$. In the particular case of NGC 4051, \citet{krongold} found $r_{sub}\simeq 0.01$ pc.} is at least one order of magnitude larger. Hence the authors suggested a model in which the black hole accretion disk is at the origin of a $X$-ray absorber wind, which forms a conical structure moving upward.
\begin{figure*}[t]
\vspace{0.4cm}
\begin{center}
$\begin{array}{c@{\hspace{0.05in}}c@{\hspace{0.05in}}c}
\epsfxsize=3.55in \epsfysize=3.55in \epsffile{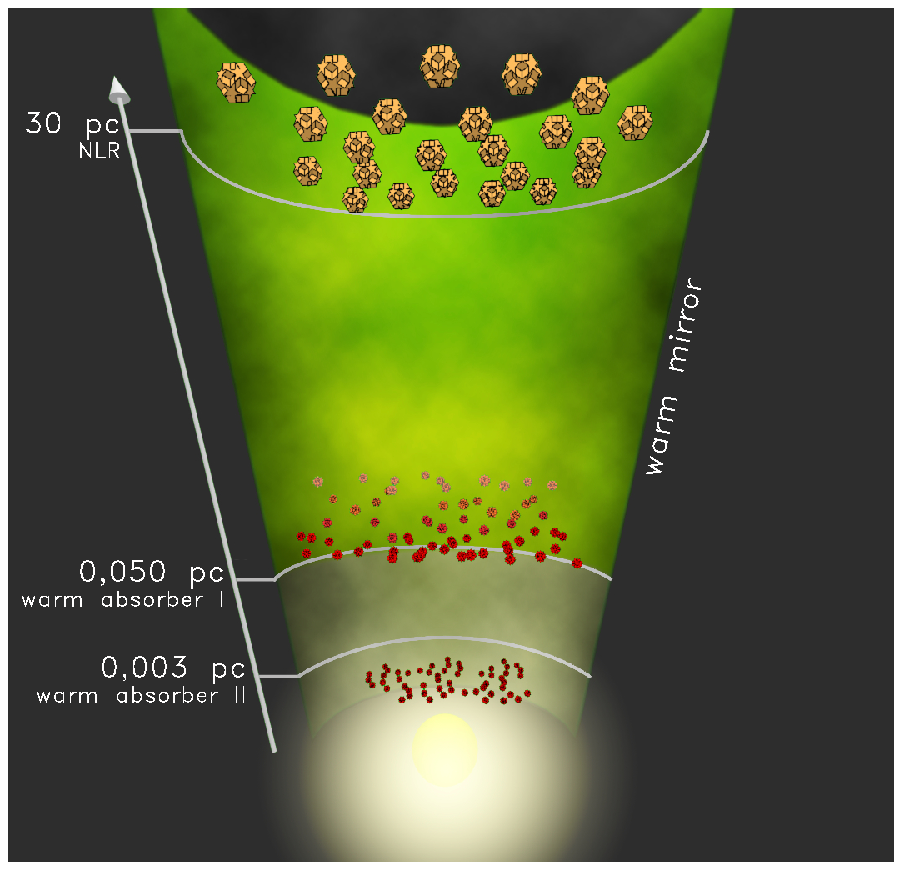} &
\epsfxsize=3.55in \epsfysize=3.55in \epsffile{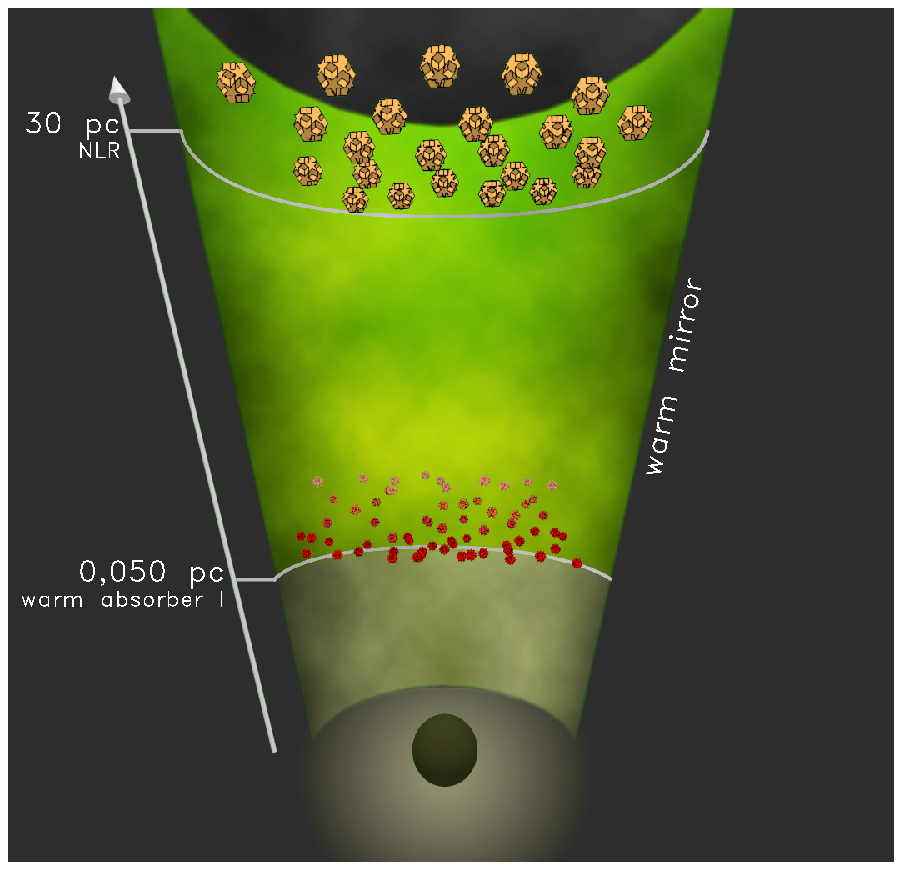}
\end{array}$
\end{center}
\caption{The cartoon shows qualitatively the location of the $X$-ray emitting and absorbing material (see text for details).}
\label{figcartoon}
\end{figure*}

If this is the correct picture, when the continuum source is switched off, the compact warm absorber might not be observed anymore during the low state flux of NGC 4051. Our analysis showed instead the existence of a more exterior $X$-ray absorber, which absorbs the soft $X$-ray photons emitted from sources (as for example the inner surface of the conical structure proposed by \citealt{krongold}) located (in projection) at scales larger than the torus and/or the narrow-line regions. Remarkably, this could indicate the existence of a diffuse warm material filling the wind-generated cone.

Figure \ref{figcartoon} gives a qualitative representation of the model.
During the high state flux (left panel) a two ionization component warm absorber (here labeled as II) lying within a few l-days ($\simeq 0.003$ pc) from the accreting black hole was identified by \citet{krongold}. \citet{warmabsorber} found that the NGC 4051 $X$-ray narrow-line regions can be placed at a distance of $r>0.02$ pc, while the optical narrow-line regions are on the scale of tenth of parsec (\citealt{christopoulou1997}). During the low state flux (right panel), the interior warm absorber might not be observed anymore since the central engine is switched off. A more exterior warm absorber (labeled as I) could now absorb the $X$-ray photons emitted from sources located on the scale larger than the torus and/or the narrow-line regions.

\begin{acknowledgements}
This paper is based on observations from {\it XMM}-Newton, an
ESA science mission with instruments and contributions directly funded by ESA
Member States and NASA. We are grateful to the anonymous referee for the suggestions that improved the paper a lot. AAN is grateful to R. Porter for help with the Cloudy code, to C. Gordon for solving a few issues with the XSPEC package and to Yair Krongold for many fruitful conversations while writing this paper. Our acknowledgements also to Marco Castelli for drawing the cartoon in Fig. 8.
\end{acknowledgements}



\begin{thebibliography}{99}
\bibitem[\protect\citeauthoryear{Armentrout et al.}{2007}]{armentrout2007}
Armentrout, B.K., Kraemer, S.B., \& Turner, T. J, 2007, ApJ, 665, 237

\bibitem[\protect\citeauthoryear{Arnaud et al.}{2007}]{arnaud}
Arnaud, K., Dorman, B., \& Gordon, C., 2007, ApJ, {\it An X-ray Spectral Fitting Package - User's Guide for version 12.4.0}, Heasarc Astrophysics Science Division

\bibitem[\protect\citeauthoryear{Bianchi et al.}{2006}]{bianchi2006}
Bianchi, S., Guainazzi, M. \& Chiaberge M., 2006, A\&A  448, 499

\bibitem[\protect\citeauthoryear{Bianchi et al.}{2010}]{bianchi2010}
Bianchi, S., Chiaberge M., Evans, A.D., Guainazzi, M., et al., 2010, MNRAS, in press

\bibitem[\protect\citeauthoryear{Blustin et al.}{2005}]{blustin2005}
Blustin, A.J., Page, M.J., Fuerst, S.V. et al., 2005, A\&A, 431, 111

\bibitem[\protect\citeauthoryear{Cash}{1979}]{cash}
Cash, W., 1979, ApJ, 228, 939

\bibitem[\protect\citeauthoryear{Collinge et al}{2001}]{collinge}
Collinge, M.J., Brandt, W. N., Kaspi, S., Crenshaw, D.M., et al., 2001, ApJ, 557, 2

\bibitem[\protect\citeauthoryear{Christopoulou et al.}{1997}]{christopoulou1997}
Christopoulou, P.E., Holloway, A. J., Steffen, W., et al., 1997, ApJ, MNRAS, 284, 385

\bibitem[\protect\citeauthoryear{Dere et al.}{2001}]{chianti}
Dere, K.P., 2001, ApJSS, 134, 331

\bibitem[\protect\citeauthoryear{Dickey \& Lockman}{1990}]{dickey}
Dickey, J.M., \& Lockman, F.J., 1990, ARA\&A, 28, 215

\bibitem[\protect\citeauthoryear{Elvis}{2000}]{elvis2000}
Elvis, M., 2000, AAS, Bulletin of the American Astronomical Society, 32, 1195

\bibitem[\protect\citeauthoryear{Ferland et al.}{1998}]{ferland}
Ferland, G.J., Korista, K.T., Verner, D.A., et al., 1998, PASP, 110, 761

\bibitem[\protect\citeauthoryear{Ferland}{2008}]{hazy1}
Ferland, G.J., 2008, Hazy 1, {\it A brief Introduction To Cloudy, Introduction And Commands},
available at http://www.nublado.org

\bibitem[\protect\citeauthoryear{Guainazzi \& Bianchi}{2007}]{bianchi2007}
Guainazzi, M. \& Bianchi, S., 2007, MNRAS, 374, 1290

\bibitem[\protect\citeauthoryear{Guainazzi et al.}{1998}]{guainazzi98}
Guainazzi, M., Nicastro, F., Fiore, F., et al., 1998, MNRAS, 301, 1

\bibitem[\protect\citeauthoryear{Guainazzi et al.}{2009}]{guainazzi2009}
Guainazzi, M., Risaliti, G., Nucita, A.A., et al., 2009, A\&A  505, 589

\bibitem[\protect\citeauthoryear{Kinkhabwala et al.}{2002}]{ali}
Kinkhabwala, A., Sako, M., Behar, E.;, et al., 2002, ApJ, 575, 732

\bibitem[\protect\citeauthoryear{Krongold et al.}{2007}]{krongold}
Krongold, Y., Nicastro, F., Elvis, M., et al., 2007, ApJ, 659, 1022

\bibitem[\protect\citeauthoryear{Korista et al.}{1997}]{korista1997}
Korista, K., Baldwin, J., Ferland, G., \& Verner, D., 1997, ApJS, 108, 401

\bibitem[\protect\citeauthoryear{Korista et al.}{1997}]{korista}
Korista, K., Ferland, G. \& Baldwin, J., 1997, ApJ, 487, 555

\bibitem[\protect\citeauthoryear{Lamer et al.}{2003}]{lamer2003}
Lamer, G., McHardy, I. M., Uttley, P. \& Jahoda, K., 2003, MNRAS, 338, 323

\bibitem[\protect\citeauthoryear{Lawrence et al.}{1987}]{lawrence1987}
Lawrence, A., Watson, M.G., Pounds, K.A. \& Elvis, M., 1987, Nature, 325, 694

\bibitem[\protect\citeauthoryear{Liedhal}{1999}]{lied}
Liedahl, D. A., 1999, in {\it X-Ray Spectroscopy in Astrophysics, Lectures held at the Astrophysics School X}, Edited by J. van Paradijs and J. A. M. Bleeker. 520, 189

\bibitem[\protect\citeauthoryear{Longinotti et al.}{2007}]{longinotti2007}
Longinotti, A.L., Sim, S. A., Nandra, K., \& Cappi, M.M., 2007, MNRAS, 374, 237

\bibitem[\protect\citeauthoryear{Longinotti et al.}{2008}]{longinotti2008}
Longinotti, A.L., Nucita, A.A., Santos, Lleo M., \& Guainazzi, M., 2008, A\&A, 484, L311

\bibitem[\protect\citeauthoryear{Longinotti et al.}{2010}]{longinotti2009}
Longinotti, A.L., Costantini, E, Petrucci, P.O., Boisson, C., et al., 2010, A\&A, 510, 92

\bibitem[\protect\citeauthoryear{McHardy et al.}{2004}]{mchardy}
McHardy, I.M., Papadakis, I.E., Uttley, P., et al., 2004, MNRAS, 348, 783

\bibitem[\protect\citeauthoryear{Ogle et al.}{2004}]{warmabsorber}
Ogle, P.M., Mason, K.O., Page, M.J., et al., 2004, ApJ, 606, 151

\bibitem[\protect\citeauthoryear{Ponti et al.}{2006}]{ponti2006}
Ponti, G., Miniutti, G., Cappi, M., et al., 2006, MNRAS, 368, 903.

\bibitem[\protect\citeauthoryear{Porquet \& Dubau}{2000}]{porquet}
Porquet, D., \& Dubau, J., 2000, A\&AS, 143, 495

\bibitem[\protect\citeauthoryear{Porter et al.}{2006}]{porter}
Porter, R.L., Ferland, G.J., Kraemer, S.B., et al., 2006, PASP, 118, 920

\bibitem[\protect\citeauthoryear{Porter \& Ferland.}{2007}]{porter2007}
Porter, R., \& Ferland, G., 2007, ApJ 664, 586

\bibitem[\protect\citeauthoryear{Pounds et al.}{2004}]{pounds}
Pounds, K.A., Reeves, J.N., king, A.R. \& Page, K.L., 2004, MNRAS, 350, 10

\bibitem[\protect\citeauthoryear{Press et al.}{2004}]{press}
Press, W.H., Teukolsky, S.A., Vetterling, W.T., \& Flannery, B.P., 1999, {\it Numerical Recipes in Fortran 77, The art of Scientific Computing}, Second Edition, Cambridge University press

\bibitem[\protect\citeauthoryear{Reynolds et al.}{1995}]{reynolds}
Reynolds, C.S., Fabian, A.C., Nandra, K., et al., 1995, MNRAS, 277, 901

\bibitem[\protect\citeauthoryear{Sako et al.}{2000 a}]{sakoa}
Sako, M., Kahn, S.M., Paerels, F., \& Liedahl, D.A., 2000a, ApJ, 542, 684

\bibitem[\protect\citeauthoryear{Sako et al.}{2000 b}]{sakob}
Sako, M., Kahn, S.M., Paerels, F., \& Liedahl, D.A., 2000b, ApJ, 543, L115

\bibitem[\protect\citeauthoryear{Sambruna et al.}{2001}]{sambruna}
Sambruna, R. M., Netzer, H., Kaspi, S., Brandt, W. N., et al., 2001, ApJ, 546, 13

\bibitem[\protect\citeauthoryear{Schmitt et al.}{2003}]{schmitt2003}
Schmitt, H.R., Donley, J.L., Antonucci, R.R.J., et al., 2003, ApJ, 597, 768

\bibitem[\protect\citeauthoryear{Steenbrugge et al.}{2009}]{steenbrugge}
Steenbrugge, K.C., Fenovcik, M., Kaastra, J.S., et al., 2009, A\&A, 496, 107

\bibitem[\protect\citeauthoryear{Tarter, Tucker \& Salpeter}{1969}]{tarter}
Tarter, C.B., Tucker, W.H., Salpeter, E.E., 1969, ApJ, 156, 943

\bibitem[\protect\citeauthoryear{Uttley et al.}{1999}]{uttley}
Uttley, P., McHardy, I.M., Papadakis, I.E., et al., 1999, Nucl. Phys. B, 69, 490

\bibitem[\protect\citeauthoryear{Uttley et al.}{2003}]{uttley2203}
Uttley, P., Fruscione, A., McHardy, I., \& Lamer, G., 2003, ApJ, 595, 656

\bibitem[\protect\citeauthoryear{{\it XMM}-Newton Users Handbook}{2009}]{xmmuserbook}
{\it XMM}-Newton Users Handbook, 2009, Issue 2.7, Edited by Ness J.-U. et al.

\end{thebibliography}
\end{document}